\documentclass[a4paper,11pt]{article}
\usepackage{pos}
\usepackage{bibspacing}
\title{Cosmic rays in the GeV-TeV energy range from two types of supernovae}

\author*[a]{Satyendra Thoudam}
\author[b]{Bj\"orn Eichmann}
\author[c]{J\"org P. Rachen}
\affiliation[a]{Department of Physics, Khalifa University, P.O. Box 127788, Abu Dhabi, United Arab Emirates}
\affiliation[b]{Ruhr Astroparticle and Plasma Physics Center (RAPP Center), Ruhr-Universit\"at Bochum\\
Institut f\"ur Theoretische Physik IV, 44780 Bochum, Germany}
\affiliation[c]{Astrophysical Institute, Vrije Universiteit Brussel (VUB), Pleinlaan 2, 1050 Brussels, Belgium}

\emailAdd{satyendra.thoudam@ku.ac.ae}

\abstract{The AMS-02 experiment has reported precise measurements of energy spectra of several cosmic-ray species in the range of ${{{\sim}}}\,(0.5-2000)$~GeV/n. An intriguing finding is the differences in the spectral shape between the different species. Protons exhibit the steepest spectrum of all the species, and helium, carbon, oxygen and iron spectra are found to be harder than that of neon, magnesium and silicon. These observations are difficult to explain as diffusive shock acceleration, the currently most plausible theory for cosmic particle acceleration at high energies, expects independence of the spectral index from mass and charge of the accelerated particle. Moreover, propagation in the Galaxy has been shown to not being able to compensate for this discrepancy. In this work, we present an explanation based on two-component model for the origin of cosmic rays in the Galaxy -- the first component originating from regular supernova remnants in the interstellar medium and the second component from Wolf-Rayet supernovae. Using recent results on cosmic-ray injection enhancement at supernova shocks in the uniform interstellar medium and in the wind environment of Wolf-Rayet stars, we show that the combination of the two components may explain most of the behavior observed by the AMS-02 experiment.}

\FullConference{37$^{\rm{th}}$ International Cosmic Ray Conference (ICRC 2021)\\
		July 12th -- 23rd, 2021\\
		Online -- Berlin, Germany}

\begin{document}
\maketitle

\section{Introduction}
The AMS-02 experiment has provided unprecedented high-precision measurements of several species of cosmic rays ranging from protons to iron nuclei as well as electrons and positrons \cite{ams-review}. Their measurements, along with those from other current generation experiments such as CREAM \cite{cream2017}, PAMELA \cite{pamela2014}, CALET\cite{calet2019} and DAMPE \cite{dampe2019}, have revealed several new properties of cosmic rays which have severely challenged our current understanding of the acceleration and propagation of these high-energy particles in the Galaxy. One such finding is the different spectral shapes between the primary cosmic-ray species -- protons exhibiting the steepest spectrum followed by neon, magnesium and silicon nuclei, and then helium, carbon, oxygen and iron nuclei showing the hardest spectra. Particle acceleration by the diffusive shock acceleration mechanism and the subsequent propagation of particles in the Galaxy predict an index that is independent of the type and charge of the nuclei \cite{drury1983}.

We present here a possible explanation of the spectral differences using a two-component model for the origin of cosmic rays in the Galaxy, 
with regular supernova remnants in the interstellar medium (SNR-CRs) being the first, and supernova explosions in the wind environment of Wolf-Rayet (WR) stars, hereafter donoted as WR-CRs, the second component \cite{thoudam2016}. The model has been 
previously 
shown to successfully explain the unexpected light composition of cosmic rays at and above the "second knee" recently observed by the LOFAR radio telescope \cite{buitink2016} and the Pierre Auger Observatory \cite{auger2014}. Here we present first results from including recent theoretical developments on element injection and non-thermal spectral behaviour into the model.

\section{Cosmic rays from regular and Wolf-Rayet supernovae}
\label{CRs}
Theoretically, it has been established that shock waves associated with supernova remnants can produce cosmic rays by accelerating particles from the thermal pool present in the ambient medium. The underlying diffusive shock acceleration process produces cosmic rays with a power-law spectrum of index close to $-2$ (see e.g., \cite{bell1978}). However, new studies indicate that the spectral shape can differ depending on the nature of the ambient medium, and that shocks in a WR wind environment may produce a harder spectrum than in the uniform interstellar medium \cite{er2021}.

Based on the recent results on non-linear diffusive shock acceleration theory \cite{hc2020}, it is expected that shocks that pass through a uniform interstellar environment form a slightly steeper spectra than what is expected from the linear approach -- even in the case of high Mach numbers. This is in principle also true for a non-uniform WR wind environment, but due to its comparatively high ratio of radiation over gas pressure, the adiabatic index yields a value smaller than $5/3$ causing a slightly harder spectral behavior than the uniform interstellar medium environment \cite{er2021}. In this paper, we will present some preliminary results to show  that a combination of the steeper cosmic-ray component from regular supernova remnants and the harder component from WR stars, together with the effect of re-acceleration described in Section \ref{transport}, can explain the observed spectra of protons and heavier nuclei reported by the AMS-02 experiment.

Another distinction between the SNR-CRs and WR-CRs is in the presence of protons. WR winds show a negligible hydrogen content \cite{pollock2005}, while the composition of the interstellar medium is expected to follow the elemental solar abundance. Hence WR-CRs are expected to be depleted of protons, and the observed cosmic-ray protons are expected to come mostly from the SNR-CR component. WR-CRs can contribute mainly in the case of helium and other heavier nuclei.

Most of the supernova explosions in our Galaxy occur in the interstellar medium. Only a small fraction is expected to occur in the winds of WR stars \cite{gal2014}. A frequency of ${{\sim}}\,1$\,WR supernova explosion in every $210$~years has been estimated \cite{thoudam2016}. For the commonly  assume supernova rate of ${{\sim}}\,1/30$~years, this corresponds to ${{{\sim}}}\,1$\,WR supernova for every $7$~supernova explosions in the Galaxy. This already suggests that the contribution of WR-CRs compared to that of SNR-CRs is expected to be small at low energies, but slowly increase with energy because of their harder source spectrum if we assume that both the types of supernovae are channeling the same fraction of kinetic energy into cosmic rays which, in reality, can be somewhat different.

Within a source class, we will consider the same source index for the different types of cosmic-ray primary species. This is unlike most other studies where the source index is taken to be different for different nuclei in contrast to the prediction from shock acceleration theory. Moreover, the source normalization will be scaled according to the thermal abundance and cosmic-ray injection efficiency at supernova shocks, but constrained within the efficiency limits provided by the simulation of particle acceleration in different environments. In most studies, the source normalization is kept as a free parameter which is arbitrarily adjusted based on the observed spectrum.

\section{Cosmic-ray transport including re-acceleration}
\label{transport}
After cosmic rays escape from the supernova remnants, they undergo diffusive propagation through the Galaxy. The propagation can be accompanied by inelastic collisions with the interstellar matter and also re-acceleration from repeated encounters with expanding supernova remnant shock waves in the interstellar medium. The re-acceleration, which is expected mostly from older remnants (so weaker shocks), can produce a bump in the cosmic-ray spectrum in the GeV region and  explain the observed spectral hardening in the proton and helium nuclei at ${{\sim}}\,300$~GeV/n \cite{thoudam2014}.

The propagation of cosmic rays in the Galaxy undergoing re-acceleration can be described by the steady-state transport equation \cite{thoudam2014},
\begin{align}
\label{eq-transport}
\nabla\cdot(D&\nabla N)-\left[\bar{n} v\sigma+\xi\right]\delta(z)N+\left[\xi sp^{-s}\int^p_{p_0}du\;N(u)u^{s-1}\right]\delta(z)=-Q\delta(z),
\end{align}
where we assume a cylindrical geometry for the propagation region with vertical halo boundaries at $z=\pm L$ and no boundaries along the radial direction $r$. We further assume the Earth to be located at the Galactic plane $z=0$. $N(r,z,p)$ represents the differential number density of cosmic rays with momentum/nucleon $p$, and $Q(r,p)\delta(z)$ is the cosmic-ray injection rate per unit volume in the Galaxy. $D(\rho)=D_0\beta(\rho/\rho_0)^a$ represents the diffusion coefficient as a function of the particle rigidity $\rho$, where $D_0$ is the diffusion constant, $\beta=v/c$ with $v(p)$ and $c$ representing the velocity of the particle and the velocity of light respectively, $\rho_0=3$~GV is a constant, and $a$ is the diffusion index. The rigidity is defined as $\rho=Apc/Ze$, where $A$ and $Z$ represent the mass number and the charge number of the nuclei respectively, and $e$ is the charge of an electron. $\bar{n}$ represents the surface density of interstellar matter in the Galactic disk, $\sigma(p)$ is the inelastic interaction cross-section, and $\xi$ corresponds to the rate of re-acceleration. We take $\xi=\eta V\bar{\nu}$, where $V=4\pi \Re^3/3$ is the volume of a supernova remnant of radius $\Re$ re-accelerating cosmic rays, $\eta$ is a correction factor that accounts for the actual unknown size of the remnants, and $\bar{\nu}$ is the frequency of supernova explosions per unit surface area in the Galactic disk, which includes the sum of regular supernovae in the interstellar medium $\bar{\nu}_\mathrm{SNR}$ and those in the WR stars environment $\bar{\nu}_\mathrm{WR}$. In Equation \ref{eq-transport}, we assume that both the cosmic-ray sources and the interstellar matter are uniformly distributed in the Galactic plane and extended up to a radius $R$.

We assume that re-acceleration instantaneously produces a power-law spectrum of index $s$. The source term $Q(r,p)$ can be expressed as $Q(r,p)=\bar{\nu}_\mathrm{s} \mathrm{H}[R-r]\mathrm{H}[p-p_0]Q(p)$, where $\bar{\nu}_\mathrm{s}$ represents either $\bar{\nu}_\mathrm{SNR}$ or $\bar{\nu}_\mathrm{WR}$ depending on the type of the source we consider, $\mathrm{H}(m)=1 (0)$ for $m>0 (<0)$ represents a Heaviside step function, and the source spectrum $Q(p)$ is assumed to follow a power-law in total momentum which, in terms of momentum/nucleon, can be written as $Q(p)=AQ_0 (Ap)^{-q}$,
where $Q_0$ is a normalization constant which is proportional to the amount of energy $f$ channelled into cosmic rays by a single supernova event, and $q$ is the spectral index. Hereafter, the parameter $q$ in Equation (2) will be denoted by $q_\mathrm{SNR}$ and $q_\mathrm{WR}$ for the regular and WR supernovae respectively. For the parameter $f (\equiv Q_0)$, we use $f_\mathrm{SNR, P}$ for SNR-CR protons and $f_\mathrm{WR, He}$ for WR-CR helium. For any heavy nuclei $X$, we take, $f_\mathrm{SNR, X}\equiv f_\mathrm{SNR, P}N_\mathrm{sol}F_\mathrm{SNR, X}$, where $N_\mathrm{sol}$ is the elemental solar abundance with respect to hydrogen, and $F_\mathrm{SNR, X}$ is an enhancement factor that accounts for the injection efficiency of a given species with respect to protons at the supernova shocks in the uniform interstellar medium. Similarly, we take, $f_\mathrm{WR, X}\equiv f_\mathrm{WR, He}N_\mathrm{wind}F_\mathrm{WR, X}$ for the case of the WR supernova, where $N_\mathrm{wind}$ is the wind composition with respect to helium and $F_\mathrm{WR, X}$ is the enhancement factor with respect to helium. These enhancement factors include the thermal-to-non-thermal element abundance ratios that have been recently explored in comparison to low energy cosmic ray data on the base of a combined injection scenario of gas ions and charged dust grains into diffusive shock acceleration \cite{EichmannRachenJCAP}.
 
For cosmic-ray secondaries such as boron nuclei, their transport equation can also be described by Equation (1), but with the source term replaced by 
\begin{equation}
Q_2(r,p)=\bar{n} v_1(p)\sigma_{12}(p)H[R-r]H[p-p_0]N_1(r,p) \delta(z),
\end{equation}
where $v_1$ is the velocity of the primary nuclei, $\sigma_{12}$ is the total fragmentation cross-section of the primary to the secondary, and $N_1$ is the density of the primary nuclei. The subscripts $1$ and $2$ have been introduced to denote the primary and secondary nuclei respectively. Solutions of Equation (1) for both the primary and secondary nuclei can be found in Ref. \cite{thoudam2014}.

\section{Method of calculation}
\label{method}
The model parameters include the propagation and re-acceleration parameters $(D_0, \delta, s, \eta)$, source parameters $(q_\mathrm{SNR}, q_\mathrm{WR})$,  $(f_\mathrm{SNR, P}, f_\mathrm{WR, He})$, $(F_\mathrm{SNR, X}, F_\mathrm{WR, X})$, and the solar modulation parameter $\phi$. Different nuclei belonging to the same source class take the same source index $q_\mathrm{SNR}$ or $q_\mathrm{WR}$. Only $F_\mathrm{SNR, X}$ and $F_\mathrm{WR, X}$ can vary between different species, but only within reasonable limits from the simulation of cosmic-ray acceleration at astrophysical shocks in the interstellar medium or in the wind-environment of WR stars \cite{er2021}. Table \ref{table-parameters} lists the ranges for the parameters that we scan in the present study. The calculation is performed iteratively by scanning over the parameters range, and checking the residuals of the predicted boron-to-carbon ratio and the spectra of the individual species with respect to the AMS-02 data as described below:\\
\textit{Step 1:} We start with an initial carbon and oxygen spectra from previous studies \cite{thoudam2016}, and calculates the boron-to-carbon ratio  based on our model by scanning over the range of diffusion and re-acceleration parameters $(D_0, \delta, s, \eta)$ as well as the solar modular parameter $\phi$. The parameters set that best fits the AMS-02 data is determined, and these values are then fixed for the following steps.\\
\textit{Step 2:} The next step calculates the proton spectrum for the range of the source parameters $(q_\mathrm{SNR},  f_\mathrm{SNR, P})$, and compares with the observed data. This step considers only the SNR-CRs as WR-CRs have a negligible proton component. This freezes $(q_\mathrm{SNR}, f_\mathrm{SNR, P})$ for the SNR-CR protons. This also automatically fixes the source index for helium and other heavier nuclei for the SNR-CR component, and only the enhancement factor $F_\mathrm{SNR, X}$ is allowed to vary in the next steps.\\
\textit{Step 3:} Helium spectrum is calculated by scanning over both the SNR-CR  and WR-CR parameters. This determines the WR-CR source parameters $(q_\mathrm{WR}, f_\mathrm{WR, He})$ and the enhancement factor $F_\mathrm{SNR, He}$ in the interstellar medium for helium nuclei. It also fixes the source indices of heavier nuclei for the WR-CR component.\\
\textit{Step 4:} For nuclei heavier than helium (carbon and oxygen in the present study), the only parameters that can vary are the enhancement factors $F_\mathrm{SNR, X}$ and  $F_\mathrm{WR, X}$. The best-fit carbon and oxygen spectra obtained in this step are used as inputs for \textit{Step 1} in the next iteration. \textit{Steps 1} to \textit{4} are repeated until all the fits are converged. 

\begin{table}
\centering
\caption{\label{table-parameters} Model parameters, their ranges, step sizes, best-fit values and type of the spectrum fitted.}
\resizebox{\textwidth}{!}
{
\begin{tabular}{c|c|c|c|c}
\hline
Parameter							&	Range			&Step size		&	Best-fit value		&Fitted spectrum\\ 
\hline
$D_0$ (cm$^2$~s$^{-1}$)				&$(1-10)\times10^{28}$	& $0.5\times 10^{28}$&	$5\times 10^{28}$	&\\
$a$								&	$(0.29-0.33)$		& $0.01$			&	$0.31$		&Boron-to-Carbon ratio\\
$s$								&	$(3.5-4.6)$		& $0.01$			&	$4.45$			&\\
$\eta$							&	$(0.3-0.6)$		& $0.1$			&	$0.4$			&\\
$\phi$ (MV)						&	$(600-710)$		& $10$			&	$690$			&\\
\hline
$q_\mathrm{SNR}; f_\mathrm{SNR, P} (10^{51}~\mathrm{ergs})$	&	$(2.2-2.4); (3-10)$	&	$0.01; 0.1$	&	$2.33; 4.8$		&Proton spectrum\\
\hline
$q_\mathrm{WR}; f_\mathrm{WR, He} (10^{51}~\mathrm{ergs})$	&	$(2.0-2.2); (0.2-0.5)$	& $0.01; 0.01$		&	$2.10; 0.28$		&Helium spectrum\\
$F_\mathrm{SNR, He}$				&	$(0.9-1.25)$		& $0.01$			&	$1.17$			&\\
\hline
$F_\mathrm{SNR, C}; F_\mathrm{WR, C}$ &	$(10.0-15.0); (0.1-1.0)$ & $0.1; 0.01$	&	$13.0; 0.28$		&Carbon spectrum\\
$F_\mathrm{SNR, O}; F_\mathrm{WR, O}$ &	$(5.0-10.0); (1.0-2.0)$ & $0.1; 0.01$	&	$8.0; 1.2$			&Oxygen spectrum\\
\hline
\end{tabular}
}
\end{table}

\section{Preliminary results: Comparison with the observed data}
\label{compare}
The inelastic interaction cross-sections and the boron production cross-sections are taken as in Ref. \cite{thoudam2016}. The halo boundary is taken as $L=5$~kpc, and the radial extent of the source and the matter distribution as $R=20$~kpc. The surface hydrogen density as $\bar{n}=5.17\times 10^{20}$~atoms~cm$^{-2}$, where an extra $10\%$  is further added to account for the helium abundance in the interstellar medium. The supernova remnant radius for re-acceleration is taken to be $\Re=100$~pc with the size correction factor $\eta$ kept as a model parameter. Each supernova explosion is assumed to release a total kinetic energy of $10^{51}$~ergs, and the supernova explosion frequency is taken as $\bar{\nu}=25$~SNe~Myr$^{-1}$~kpc$^{-2}$. This corresponds to a rate of ${{{\sim}}}\,3$ supernova explosions per century in the Galaxy. The elemental solar abundance is taken from Ref. \cite{lodders2019}, and the WR wind composition from Ref. \cite{pollock2005}.

\begin{figure*}
\centering
\includegraphics*[width=0.56\columnwidth,angle=0,clip]{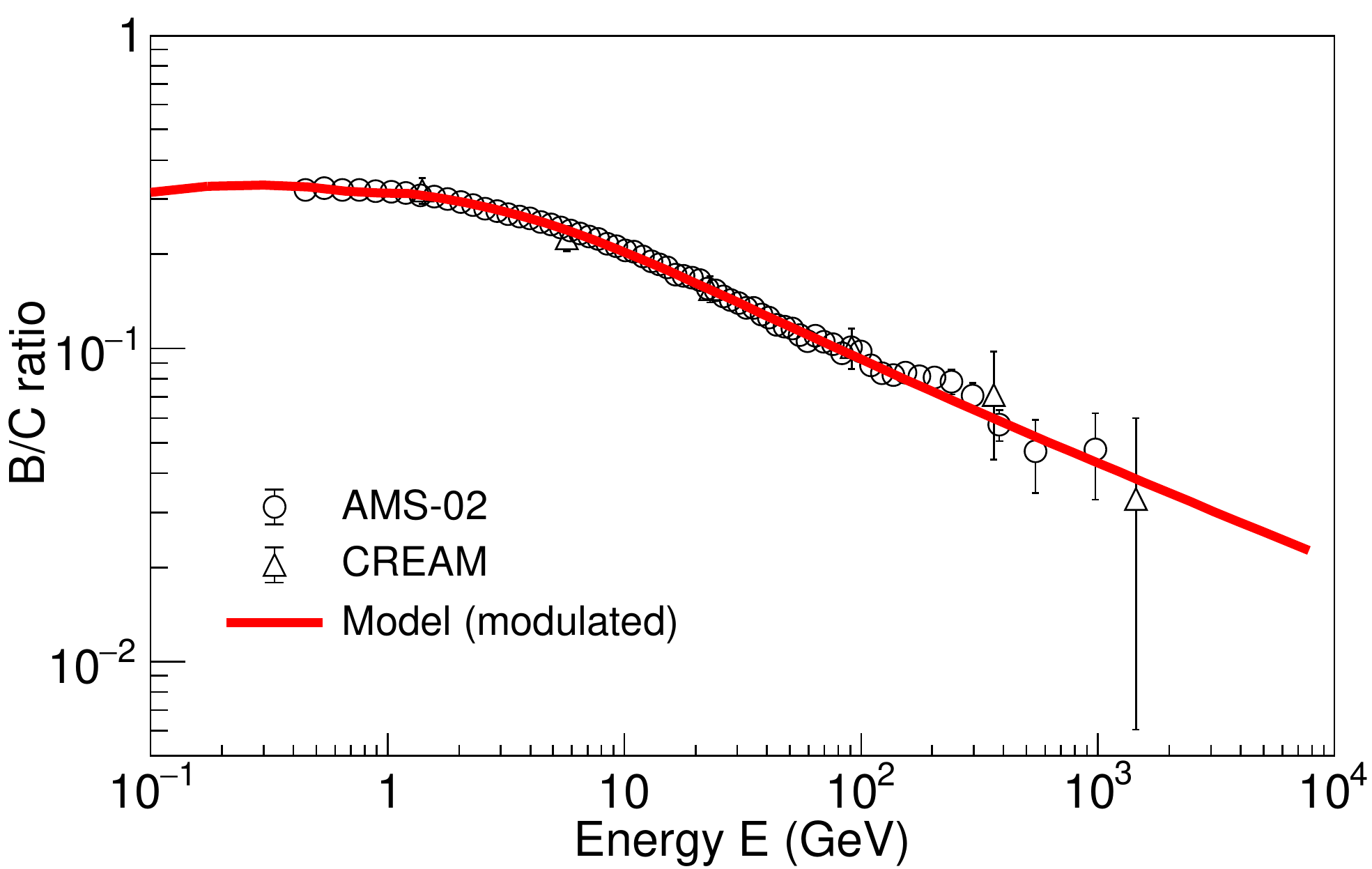}
\caption{\label {fig-BC} Boron-to-carbon ratio. \textit{Solid line:} Model prediction. \textit{Data:} AMS-02 \cite{aguilar2016}, CREAM \citep{ahn2008}. The fit is performed only on the AMS-02 data. CREAM data is shown only for comparison with the model prediction. \textit{Best-fit parameters:} $D_0=5\times 10^{28}$~cm$^2$~s$^{-1}$, $a=0.31$, $\eta=0.4$, $s=4.45$, and $\phi=690$~MV.}
\end{figure*}

It may be mentioned again that all the results presented here are obtained based on the AMS-02 data, and we show the CREAM and VOYAGER  data only for comparison with the model prediction. Result on the boron-to-carbon ratio is shown in Figure \ref{fig-BC} along with the observed data. The best-fit propagation and re-acceleration are found to be $D_0=5\times 10^{28}$~cm$^2$~s$^{-1}$, $a=0.31$, $\eta=0.4$, $s=4.45$, and the solar modulation parameters $\phi=690$~MV. 

\begin{figure*}[tbp]
\centering
\includegraphics*[width=0.49\columnwidth,angle=0,clip]{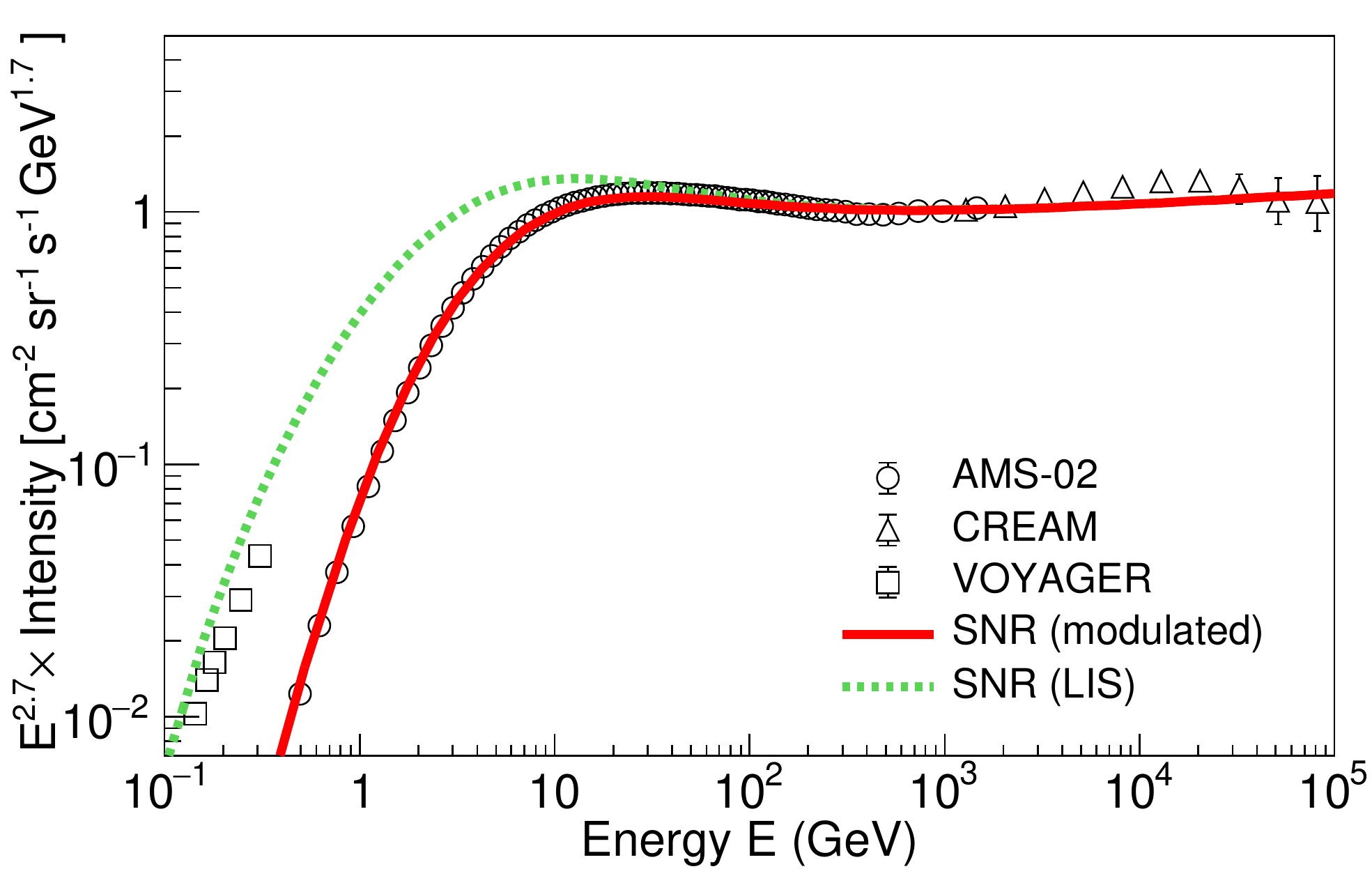}\hspace{\fill}
\includegraphics*[width=0.49\columnwidth,angle=0,clip]{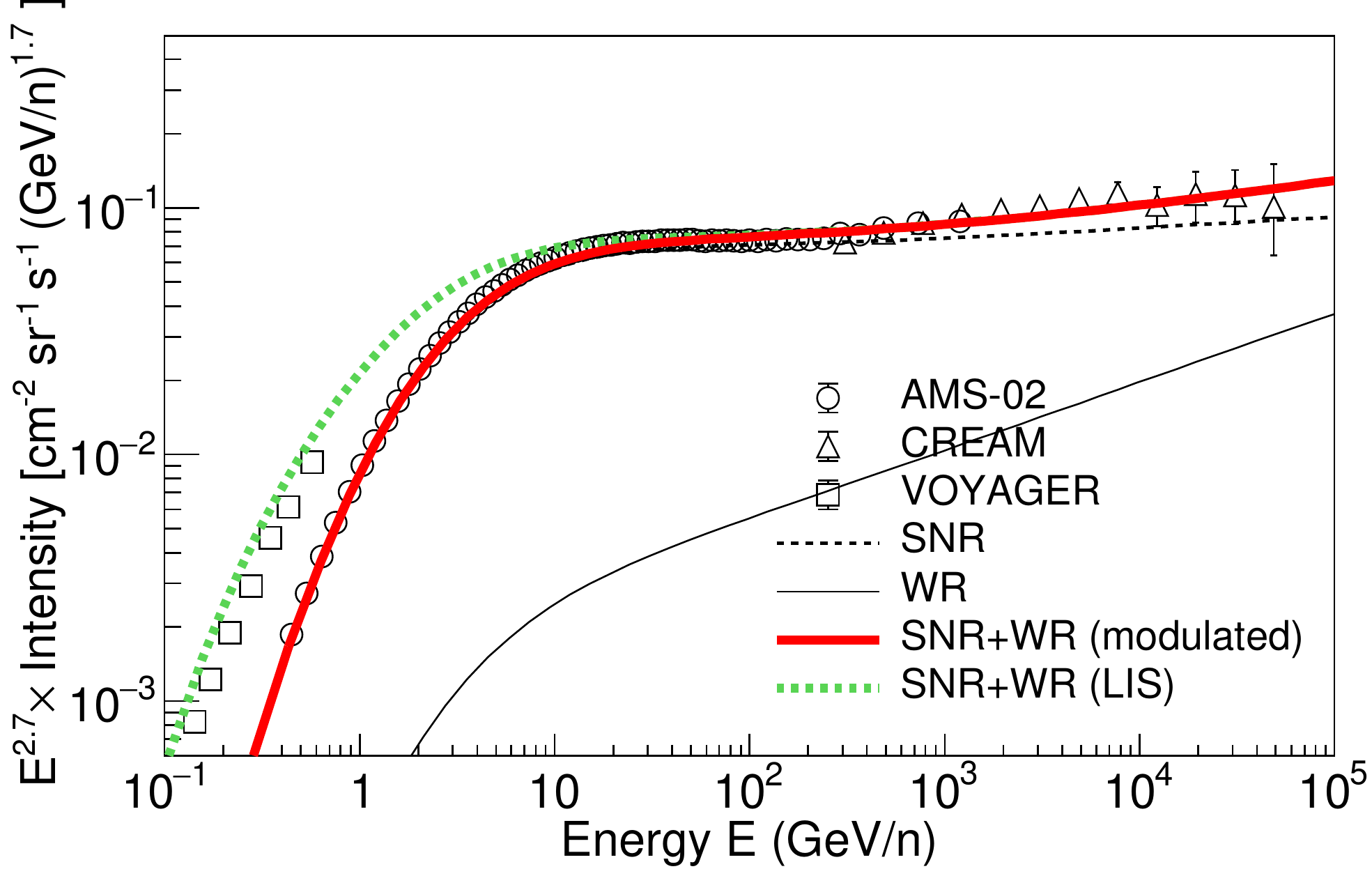}\\
\caption{\label {fig-proton-helium} Proton (left panel) and helium (right panel) spectra. \textit{Red-solid line:} Predicted total spectrum after solar modulation (for protons, same as that from   SNR-CRs). \textit{Green-dashed line:} Predicted total spectrum in the local interstellar medium. \textit{Thin-solid line:} WR-CRs contribution. \textit{Thin-dashed line:} SNR-CRs contribution. \textit{Data:} AMS-02 \cite{aguilar-p,aguilar-he}, CREAM \cite{cream2017}, VOYAGER \cite{cummings2016}. For helium, the CREAM data are  shifted by $-5\%$ in energy. The fit is performed only on the AMS-02 data. CREAM and VOYAGER data is shown only for comparison with the model predictions. \textit{Best-fit parameters:} $(q_\mathrm{SNR}=2.33,  f_\mathrm{SNR, P}=4.8\%\times 10^{51}~\mathrm{ergs})$ for protons, and $(q_\mathrm{WR}=2.10, f_\mathrm{WR, He}=0.28\%\times 10^{51}~\mathrm{ergs}, F_\mathrm{SNR, He}=1.17)$ for helium.}
\end{figure*}

Results on the proton and helium spectra are shown in Figure \ref{fig-proton-helium}. The best-fit source parameters are found to be $(q_\mathrm{SNR}=2.33,  f_\mathrm{SNR, P}=4.8\%\times 10^{51}~\mathrm{ergs})$ for protons and $(q_\mathrm{WR}=2.10, f_\mathrm{WR, He}=0.28\%\times 10^{51}~\mathrm{ergs}, F_\mathrm{SNR, He}=1.17)$ for helium nuclei. For protons, as noted earlier, the contribution comes only from the SNR-CRs, and the spectral hardening at ${{\sim}}\,300$~GeV is due to effect of re-acceleration producing a bump below ${{\sim}}\,100$~GeV [?]. For helium nuclei, SNR-CR component alone cannot satisfactorily explain the full range of the AMS-02 data as it underpredicts the data around $1$~TeV/n. Adding WR-CR component provides a better fit because of its increasing contribution at higher energies due to the flatter spectrum. For both protons and helium nuclei, the final predicted spectra (red-solid lines in Figure \ref{fig-proton-helium}) are found to agree quite well also with the CREAM data\footnote{For helium, the data points have been shifted by -5\% in energy in order to remove the systematic offset with respect to the AMS-02 data.} at higher energies although their data are not included in the fitting procedure. At low energies, the predicted spectra in the local interstellar medium (green-dashed lines in Figure \ref{fig-proton-helium}) slightly overpredicts the VOYAGER data. Future tasks will involve fine tuning the model parameters by including VOYAGER data in the fit. 

Compared to protons, the effect of re-acceleration is lower in the case of helium (or any heavier nuclei in general) because of the increasing effect of inelastic losses. But, that is not the reason for SNR-CRs for being unable to fully explain the helium data. In fact, had the re-acceleration effect been larger, the flux below ${{\sim}}\,100$~GeV/n would increase, and the predicted SNR-CRs spectrum would show even more deficit at higher energies. The observed helium spectrum is found to be too flat for a single cosmic-ray component to explain using the same source index as that of the protons. Adding the WR-CR component provides a natural explanation.

\begin{figure*}[tbp]
\centering
\includegraphics*[width=0.49\columnwidth,angle=0,clip]{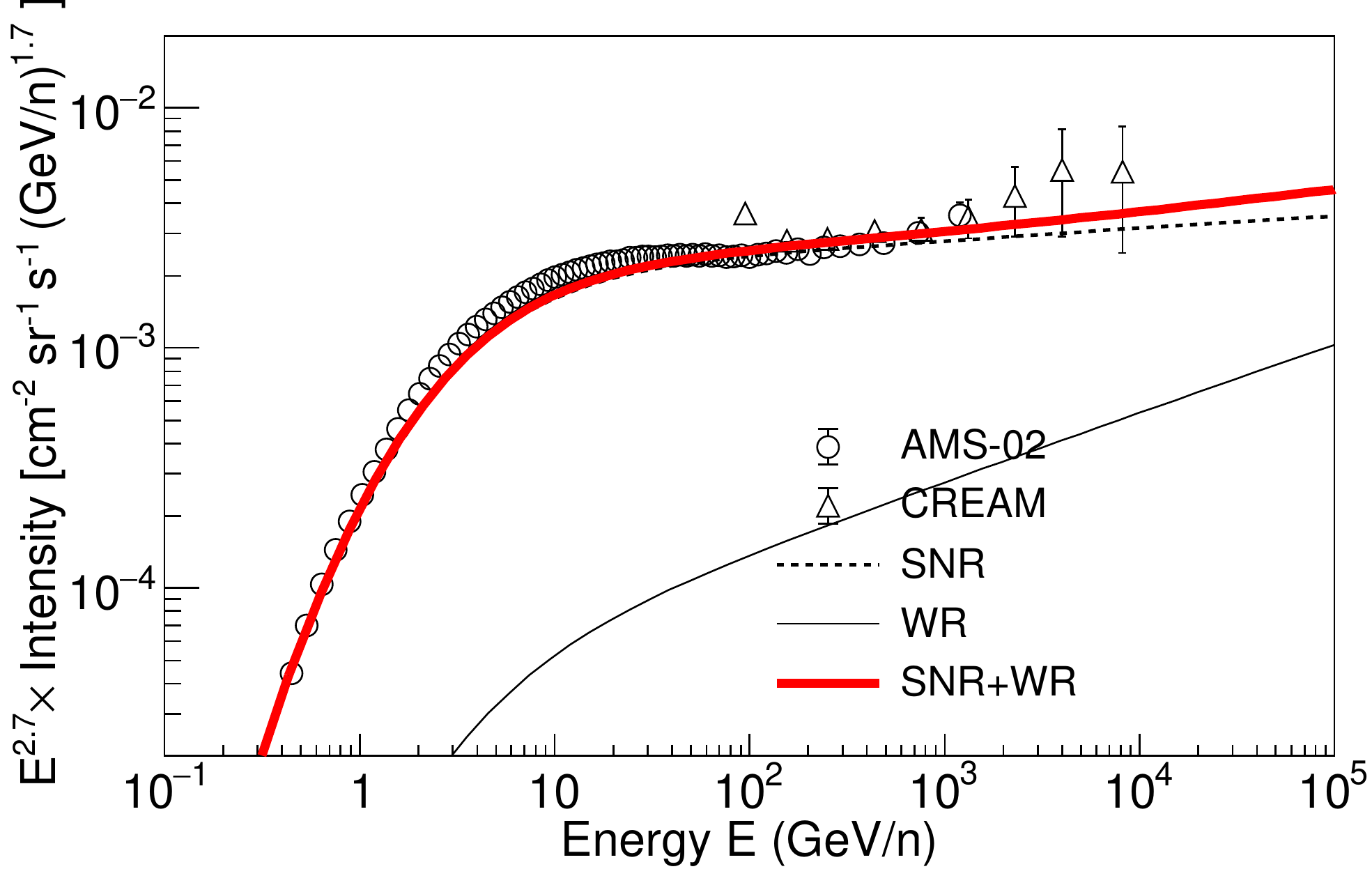}\hspace{\fill}%
\includegraphics*[width=0.49\columnwidth,angle=0,clip]{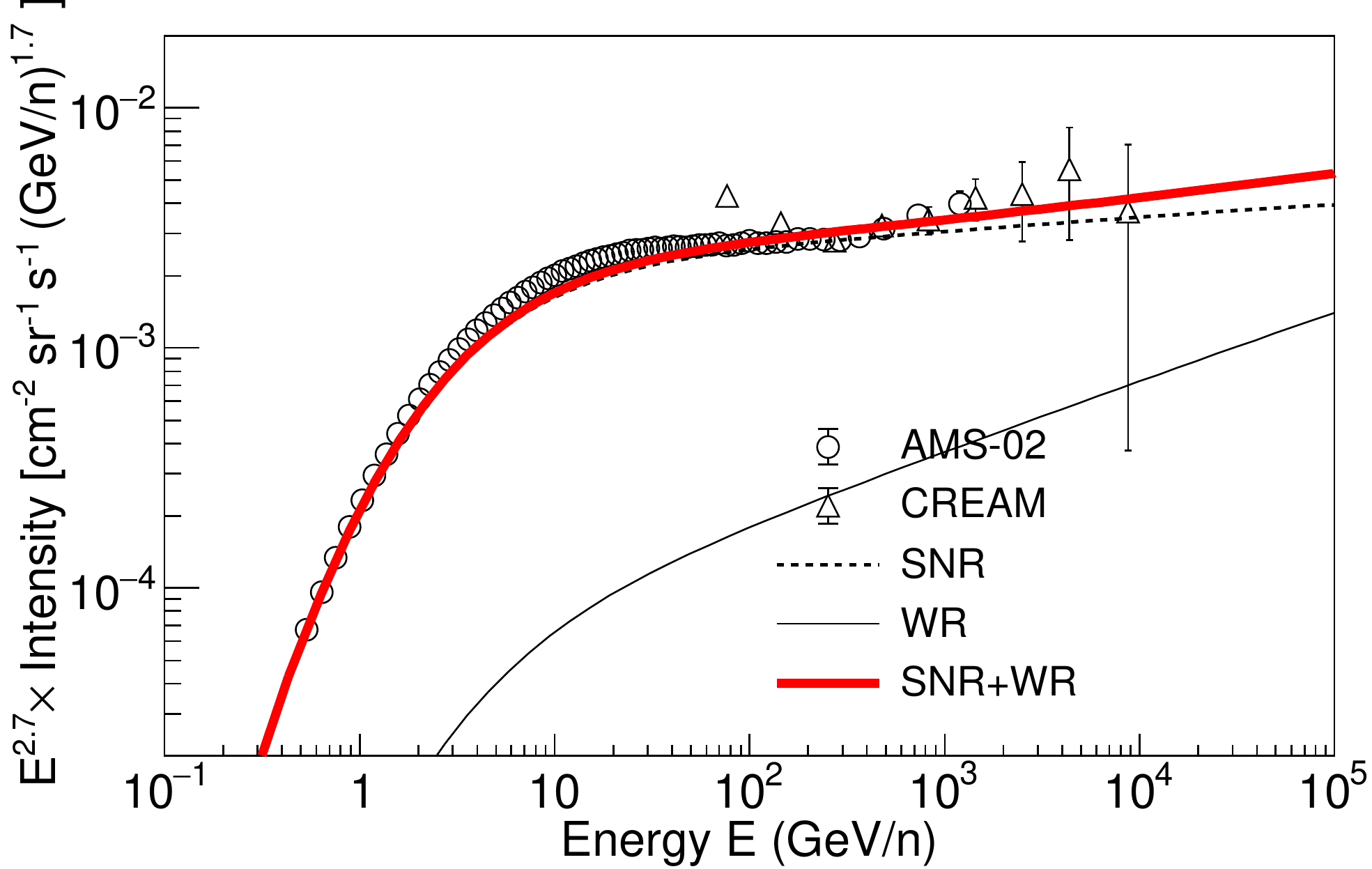}\hspace{\fill}\\
\caption{\label {fig-carbon-oxygen} Carbon (left panel) and oxygen (right panel) spectra. Different lines bear the same representation as in Figure \ref{fig-proton-helium}. \textit{Data:} AMS-02 \cite{aguilar-c}, CREAM \cite{ahn-c}. CREAM data are shifted in energy by $+10\%$ for carbon and $+20\%$ for oxygen. The fit is performed only on the AMS-02 data. CREAM data is shown only for comparison with the model prediction. \textit{Best-fit parameters:} $(F_\mathrm{SNR, C}=13.0, F_\mathrm{WR, C}=0.28)$ from the carbon fit, and $(F_\mathrm{SNR, O}=8.0, F_\mathrm{WR, O}=1.2)$ from the oxygen fit. See texts for details.}
\end{figure*}

Results for carbon and oxygen nuclei are shown in Figure \ref{fig-carbon-oxygen}. Best-fit parameters are $(F_\mathrm{SNR, C}=13.0, F_\mathrm{WR, C}=0.28)$ for carbon nuclei and $(F_\mathrm{SNR, O}=8.0, F_\mathrm{WR, O}=1.2)$. Although the predicted total spectrum (red-solid line in the figures) seems to reproduce the AMS-02 data overall, the slightly poorer fit in the region of ${{\sim}}\,(3{-}30)$~GeV/n needs some fine adjustment of the model parameters. Taking a slightly steeper source index for the SNR-CRs and a slightly harder one for the WR-CRs can improve the fit, but these will have consequences on the results of proton and helium spectra. There can also be effects from the uncertainties in the inelastic interaction cross-sections. In Figure \ref{fig-carbon-oxygen}, CREAM data are scaled in energy by $+10\%$ for carbon and $+20\%$ for oxygen.

\vspace{-3.142pt}
\section{Discussion}
\vspace{-3.142pt}
\label{discussion}
We have demonstrated that the observed spectral differences of cosmic-ray species (at least from protons to oxygen) revealed by the AMS-02 and other current experiments can be explained using cosmic rays from two different types of supernovae in the Galaxy -- regular supernova in the interstellar medium and supernova in WR stars environment. Keeping the source index same for the different species within a source class as expected from diffusive shock acceleration theory, we have found that regular supernova requires a cosmic-ray source index of $2.33$ while WR supernova requires a value of $2.1$ in order to explain the observed spectra. The flatter index for the WR component (by a difference of 0.23) agrees with the prediction from simulation of particle acceleration at astrophysical shocks in the uniform interstellar medium and in the wind environment of WR stars. In addition, the values of the cosmic-ray injection enhancement factor (see Table \ref{table-parameters}) we obtain for the uniform interstellar medium and the wind environment for the different nuclei falls within the range predicted by simulation. This enhancement factor is the only parameter that determines the spectral normalization of heavier nuclei relative to protons or helium nuclei. The results presented here has further substantiated our view of WR supernovae as the additional sources of cosmic rays in the Galaxy \cite{thoudam2016}.  
Future work will focus on optimizing the model parameters and applying the model also to other heavier species such as neon, magnesium, silicon and iron nuclei.

\noindent{\textit{Acknowledgement:} ST acknowledges funding from Khalifa University Startup grant FSU-2020-13.}
\vspace{-0.5cm}
\begin{small}
\setlength{\bibitemsep}{0.15\baselineskip}

\end{small}

%
%
%


\begin{thebibliography}{99}
\bibitem{ams-review}
Li, Z. \& Feng, J., 2021, Modern Physics Letters A, 36, 2130011
\bibitem{cream2017}
Yoon, Y.\,S., et al. 2017, ApJ, 839, 5
\bibitem{pamela2014}
Adriani, O., et al. 2014, Physics Reports, 544, 323
\bibitem{calet2019}
Adriani, O., et al. 2019, PRL, 122, 181102
\bibitem{dampe2019}
An, Q., et al., 2019, Science Advances 5, no. 9, eaax3793
\bibitem{drury1983}
Drury, L.\,O'C. 1983, Rep. Prog. Phys., 46, 973
\bibitem{thoudam2016}
Thoudam, S., et al., 2016, A\& A, 595, A33.
\bibitem{buitink2016}
Buitink, S., et al. 2016, Nature, 531, 70
\bibitem{auger2014}
Aab, A., et al., 2014, PRD, 90 122006
\bibitem{bell1978}
Bell, A.\,R. 1978, MNRAS 182, 147
\bibitem{er2021}
Eichmann, B. \& Rachen, J.\,P., PoS (ICRC2021) 466
\bibitem{hc2020}
Haggerty, C. C., \& Caprioli, D., 2020, ApJ, 905, 1
\bibitem{pollock2005}
Pollock, A.\,M.\,T., et al. 2005, ApJ, 629, 482
\bibitem{gal2014}
Gal-Yam, A., et al. 2014, Nature, 509, 471
\bibitem{EichmannRachenJCAP}
Eichmann, B. \& Rachen, J.\,P., JCAP 01 (2021) 049.
\bibitem{thoudam2014}
Thoudam, S., \& H\"orandel, J.~R. 2014, A\&A, 567, A33
\bibitem{lodders2019}
Lodders, K. 2020, Solar Elemental Abundances, Oxford University Press
\bibitem{aguilar2016}
Aguilar et al., 2016, PRL, 117, 231102
\bibitem{ahn2008}
Ahn, H.\,S., Allison, P.\,S., Bagliesi, M.\,G., et al. 2008, Astropart. Phys., 30, 133
\bibitem{aguilar-p}
Aguilar, M. et al., 2015, PRL, 114, 171103
\bibitem{aguilar-he}
Aguilar, M. et al., 2015, PRL, 115, 211101
\bibitem{cummings2016}
Cummings, A. C., et al., 2016, ApJ, 831, 18
\bibitem{aguilar-c}
Aguilar, M. et al. 2017, PRL, 119, 251101
\bibitem{ahn-c}
Ahn, H.\,S., Allison, P.\,S., Bagliesi, M.\,G., et al. 2009, ApJ, 707, 593

\end{thebibliography}
\end{document}